\documentclass{article}
\usepackage{amsmath,amssymb}
\usepackage[dvips]{graphicx}

\usepackage{latexsym}

\begin{document}

\pagestyle{plain} 
\setcounter{page}{1}
\setlength{\textheight}{700pt}
\setlength{\topmargin}{-40pt}
\setlength{\headheight}{0pt}
\setlength{\marginparwidth}{-10pt}
\setlength{\textwidth}{20cm}

\title{Generalized Clustering Coefficients and Milgram Condition for q-th Degrees of Separation  }
\author{Norihito Toyota   \and Hokkaido Information University, Ebetsu, Nisinopporo 59-2, Japan \and email :toyota@do-johodai.ac.jp }
\date{}
\maketitle

\begin{abstract}
We introduce a series of generalized clustering coefficients  based on String formalism given by Aoyama, using adjacent matrix in networks. We numerically evaluate Milgram condition proposed in order to explore q-th degrees of separation in scale free networks and small world networks.  We find that scale free network with exponent 3 just shows 6-degrees of separation. Moreover we find some relations between separation numbers and generalized clustering coefficient in both networks.
  \end{abstract}
\begin{flushleft}
\textbf{keywords:}
scale free networks, small-world networks, generalized clustering coefficient, six degrees of separation
\end{flushleft}

\section{Introduction}\label{intro}

\hspace{5mm} "Six degrees of separation" is an attractive subject that was triggered by Milgram \cite{Milg} half a century ago. Six degrees of separation indicates that people have a narrow circle of acquaintances. A series of social experiments \cite{Milg2} made the suggestion, in which all people in USA are connected through about 6 intermediate acquaintances, more certain.
    
We attack this subject based on the string formalism by fusing it into adjacent matrix description, especially how cycle   structures of networks affect the separation number $q$ \cite{Newm21}. One characteristic measure for analyzing cycle structures is the clustering coefficient \cite{Watt1}. But it only can reflect triangular structures in a network. There are many cycles with a lot of nodes. The usual clustering coefficient cannot analyses such structures. So we must introduce some indices in order to analyze more diverse cycle structures. 

 We found that the formulation which we reformulated by fusing string formalism and adjacent matrix straightforwardly makes the generalization of usual (3-th) clustering coefficient that reflects only triangle structures in a network \cite{Toyota3}. So we could analyze the effects of cycles with any nodes on the separation number by evaluating the generalized clustering coefficients.  
 
The crucial point of the reformulation is to count the number of non-degenerate strings \cite{Toyota3}, which are defined as subgraphs without  any multiple edges and/or any cycles in its subgraphs. In this paper we first give a systematic way for it and give the expressions of a series of p-th generalized clustering coefficient based on the string formalism. Secondly we evaluate Milgram condition proposed by Aoyama \cite{Aoyama}, to 
analyze the separation number on networks by using the reformulation for scale free networks \cite{Albe2} and small-world networks \cite{Watt1}. Lastly we discuss the relations between the number of cycle and the separation number, and obtain some mathematical relations between them.  

\section{STRING FORMALISM AND ADJACENT MATRIX}
\hspace{5mm}We here review the formalism given in \cite{Toyota3}. We consider a string-like part of a graph with connected $j$ nodes and call it "$j$-$string$". Let $S_j$ be the number of non-degenerate $j$-$string$ on graph. We also define strings homeomorphic to a circle as a non-degenerate string. It is generally so difficult to calculate $S_j$ and would be impossible to calculate $S_j$ with $j>7$ at present \cite{Aoyama}.  

We reformulate the string formalism by utilizing an adjacent matrix $A=(a_{ij})$. Generally the powers, $A^2, A^3, A^4 \cdots,$ of $A$ give information as to respecting that a node connects other nodes through $2, 3, 4, \cdots$intermediation edges, respectively. The information of the connectivity between two nodes in $A^n$ also contains multiplicity of edges, generally. For resolving the degeneracy, we introduce new series of matrices $R^n$ which give information as to the connectivity of two nodes, $i_0$ and in, through $n$ intermediation edges without multiplicity. We could find that  $R^n$  is given by the following formula \cite{Toyota3}; 
\begin{equation}
[ R^n] _{i_0i_n}=\displaystyle \sum_{i_1,\cdots,i_{n-1}} a_{i_0i_1} a_{i_1i_2}\cdots a_{{i-1},i_{n}} 
\frac{\displaystyle\prod_{i_k,i_j,i_k-i_j>1}^{n}(1-\delta_{i_ki_j})}{(1-\delta_{i_0i_n})}.
\end{equation}
Here the product of the Kronecker delta in the numerator plays role of protecting of non-degeneracies and in the denominator is needed to keep closed strings, respectively. (1) makes one count Sj in a unified way. We gave the  expressions of $R^2\sim R^6$ in  \cite{Toyota3} but the expressions are too complex,  needed about 2 or 3 pages in A4 size, to describe them in this paper. 

By using its formulation, we can define the usual clustering coefficient which essentially counts the number of triangular in a network. Although there are some definitions of the clustering coefficient \cite{Newm21}, \cite{Watt1}, we adopt the usual global clustering coefficient $C(3)$ \cite{Newm21} defined by 
\begin{equation}
C_{(3)}=\frac{6\times \;number \;of \;triangles }{number \;of \;connected \;triplets }=\frac{ 6\Delta_3  }{\bar{S}_3},  
\end{equation}
where $\Delta_q$ is generally the number of polygons with $q$ edges in a network. But we need more indices in order to uncover properties of general polygon structures in a network. From (2), we can generalize it to $p-th$ generalized clustering coefficient $C(p)$ straightforwardly;  
\begin{equation}
C_{(p)}=\frac{2p\times \;number \;of \;polygons }{number \;of\; connected \;p\mbox{-}plets }=\frac{ 2p\Delta_p  }{\bar{S}_p}.
\end{equation}

$C(p)$ can be systematically evaluated by using  Rn as shown the  following relations; 
 \begin{equation}\bar{S}_p=\sum_{i,j} (R^{p-1})_{ij}/2,\end{equation}
\begin{equation}
C_{(p)}=\frac{\mbox{Tr} R^p  }{ \displaystyle \sum_{i,j}R^{p-1}}.  
\end{equation}
As an example, we obtain the following formula for the usual clustering coefficient. 
\begin{equation}
[R^2]_{ij }= [A^2]_{if} - [A^2]_{ii} \delta_{if}, \;\;\; ||A||\equiv\sum_{ij} A_{ij},   
\end{equation}
\begin{equation}\displaystyle 
C_{(3)}=\frac{\mbox{Tr} R^3  }{ \displaystyle \sum_{i,j} (A^2)_{ij}-(A^2)_{ij} \delta_{ij}  }=\frac{ \mbox{Tr} A^3  }{ 
 ||A||-\mbox{Tr} A^2 }.
\end{equation}

\section{APPLICATION TO SIX DEGREES OF SEPARATION }
\label{usage}
\hspace{5mm}We analyze general $q$-$th$ dedrees of separation, according to its formalism. For $q$-$th$ degrees of separation, Aoyama proposed a condition, so-called Milgram condition at network size $N$ \cite{Aoyama}; 
\begin{equation}
M_{q} \equiv \frac{\bar{S}_{q}}{N} \sim O(N).
\end{equation}    
This condition means that the number of $q$-strings per node is nearly equal to the size of the considering network and gives a boundary whether $q$-$th$  degrees of separation is satisfied or not. As $M_q$ grows larger, $q$-$th$ degrees of separation is easier to be satisfied. We place the focus on scale free networks with the degree distribution of $P(k) \sim k^\gamma$ and small-world networks with rewiring ratio $\alpha$ where we use Newman-Watts model  \cite{Newm22} for  the construction of small-world networks. 

Fig.1 and Fig.2 show the relation between $\log_{10} M_{q}/N $and $q$ for scale free networks with diverse $\gamma$'s and for small-world networks with $0\leq \alpha \leq 1$ at $N=200$. The interior of rectangles in both figures shows the region where (8) is satisfied. In Fig.1 $\log_{10}M_q/N $ increases linearly with $q$. We see the four degrees of separation in networks with $\gamma <2.5$ but do not recognize that nodes are linked together for $\gamma \geq 3.5$ up to six degrees of separation.$\gamma=2.75$ shows five degrees of separation and $\gamma= 3.0$, which many real-world networks have about this value of exponent, just shows six degrees of separation. This result is also justified from analyses of $A^n$ \cite{Toyota3},\cite{ToSa1}. 

We observe 4 stripes in Fig.2, which corresponds $q=2$ (below curve)$\sim q=4$ (above one), respectively. Six degrees of separation is mostly observed even at small $\alpha>0$ but nodes almost do not connect each other in 2 steps, even if $\alpha=1$ (random graph). This is consistent with the fact that the average path length rapidly decreases with $\alpha$ in small-world networks.

\section{	MILGRAM CONDITIONONDITION AND GENERALIZED     CLUSTERING  COEFFICIENT} 
\hspace{5mm}We explore the relation between Milgram condition and the generalized clustering coefficients. By this, we analyze how cycle structure on a network is related with the separation number $q$. We first define the following two quantities;
\begin{eqnarray}
X &\equiv& \sum_{p=3}^{q} C_{(p)},\\  
Y &\equiv& \log_{10} M_q.
\end{eqnarray}

 \begin{figure}[t]
\begin{center}
\includegraphics[scale=0.9,clip]{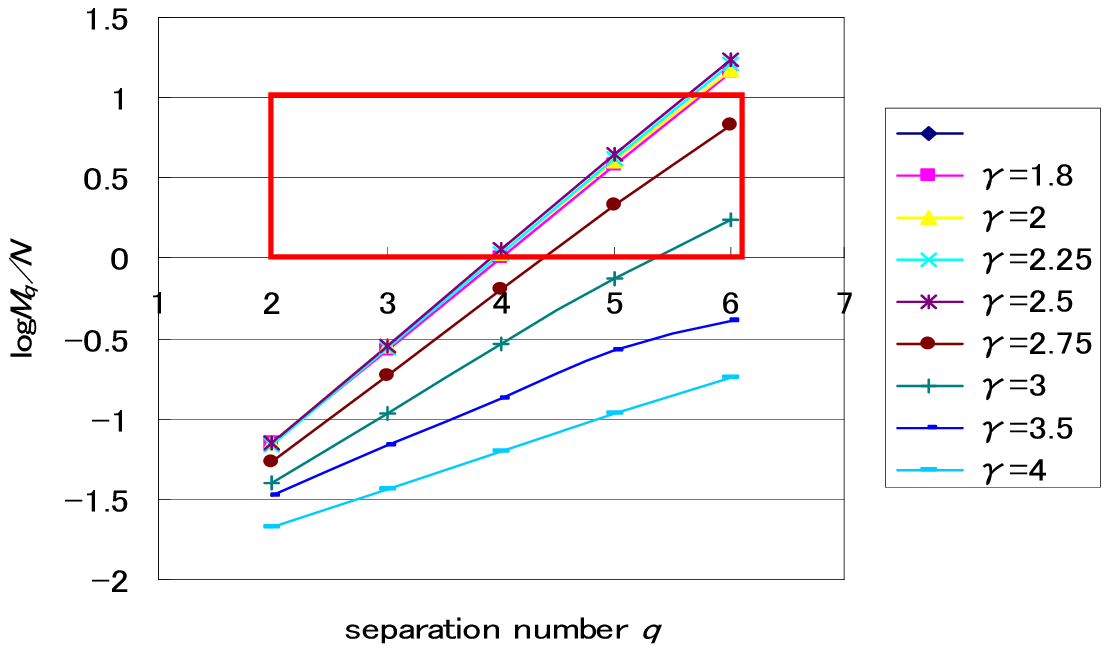} 
\caption{Separation number $q$  v.s. $\log M_q/N$ for scale free networks with  several $\gamma$.  }
\includegraphics[scale=0.9,clip]{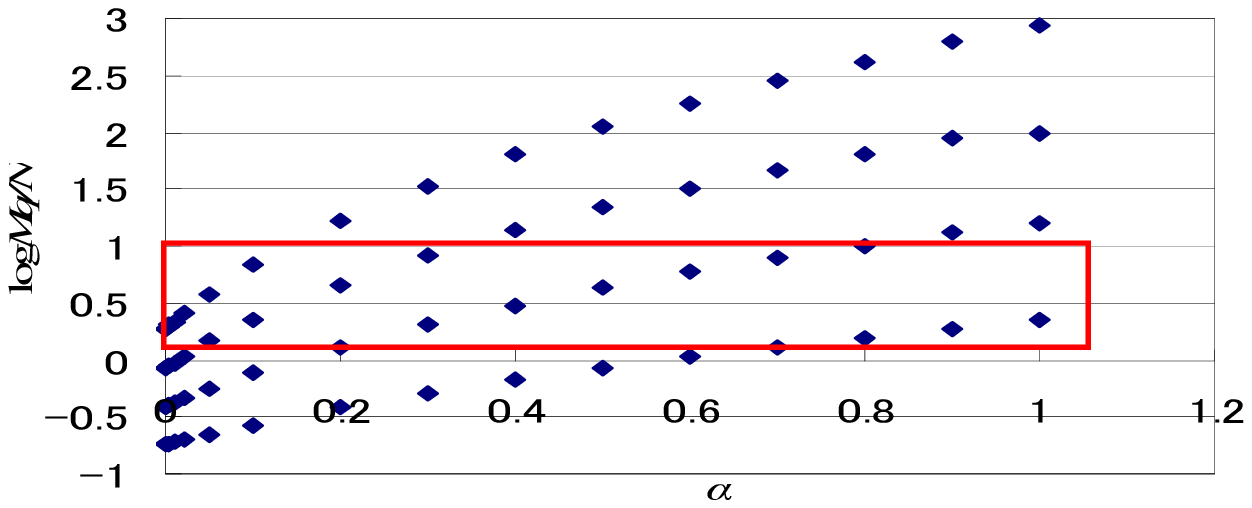} 
\end{center}
\caption{Separation number $q$ v.s.$ \log M_q/N$ for small-world networks with  several $\alpha$  }
 \end{figure} 
 
 Fig.3 and Fig.4 show the relation between $X$ and $Y$ in scale free networks with $1.8<\gamma<4.0$ and small-world networks with $0\leq \alpha \leq 1$ at $N=200$. In Fig.3, we recognize that $Y$ increases linearly with $X; \;Y =AX +B$. The two constants $A$ and $B$ are determined by numerical simulations and the values are described in Fig. 3. 
 
On the other hand, we observe 4 stripes structure in Fig.4. The meaning of four stripes is as same as Fig.2.  In this cases we find that $Y$ is a logarithmic function of $X$ by least squares method with the square of multiple correlation coefficient $R^2 >0.995$, so the relation $Y =D \log X + E$ where $D$ is a negative number holds for each $q$. The parameters $D$ and $E$ are also determined by numerical simulations but the details of the   values are not described, since they are not crucial in this discussion.
\begin{figure}[t]
\begin{center}
\includegraphics[scale=0.8,width=8cm,height=5cm,clip]{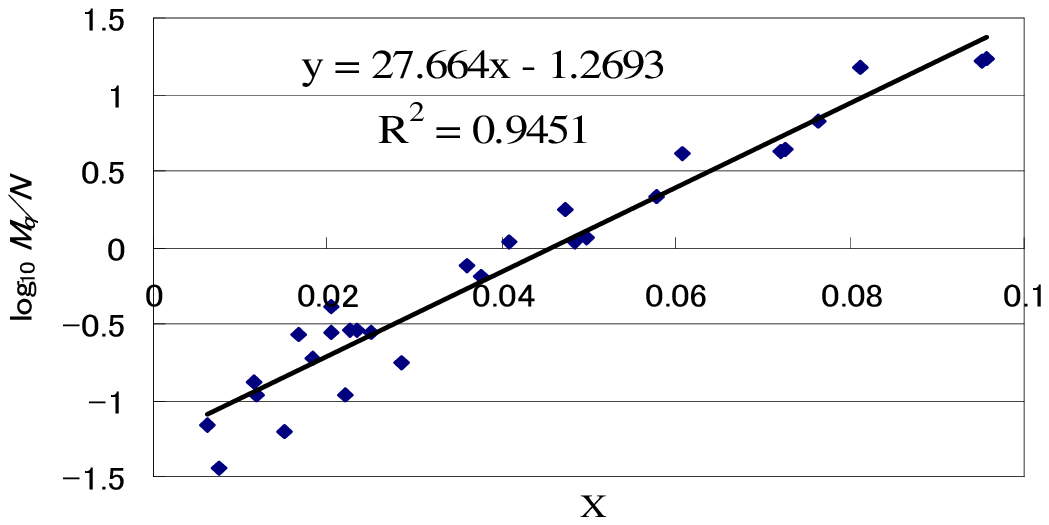}
\caption{ $X$  v.s. $\log M_q/N $for scale free networks}
\includegraphics[scale=0.7,width=8cm,height=5cm,clip]{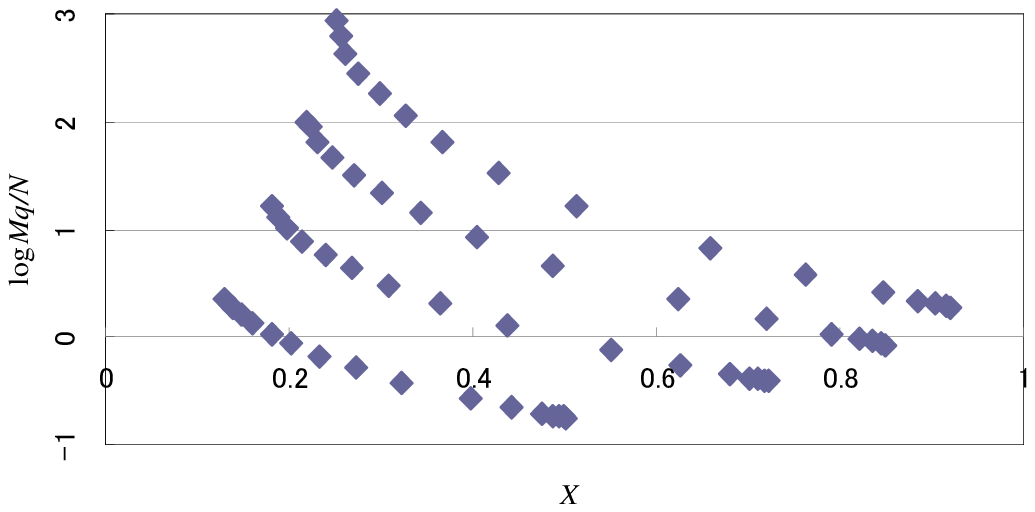}
\caption{ $X$  v.s.$ \log M_q/N$ for small-world networks}
\end{center}
\end{figure} 
So it is noticed that there are the relations of an exponential function for scale free networks and an power low for small-world networks between $\log_{10}M_q/N$ and the sum of the generalized clustering coefficient, respectively; 
\begin{eqnarray}
M_q &\sim& \exp ( c \sum_{p=3}^{q} C_{(p)}) \;\;\;\;\mbox{ for Scale free networks}, \\
M_q &\sim&  (  \sum_{p=3}^{q} C_{(p)})^{-d} \;\;\;\;\mbox{ for Small-world networks}, 
\end{eqnarray}  
where c and d are some constants determined by $A$ and $B$ and determined by $D$ and $F$, respectively. This means that the effect of $C(p)$, that is to say, cycle structures,  has a greater influence on the scale free networks than small-world ones. In scale free networks, a crucial factor that decreases the separation number is hubs and the number of cycles is actually very few and cycles only appear incidentally. So it is considered that the effect of $C(p)$ corresponding to short cuts on Milgram condition would be relatively serious.  On the other hand, a crucial factor that decreases the separation number is short cuts, which do not drastically affect cycle structures consist of few nodes in Newman-Watts model, in small-world networks. Moreover an increase in $C(p)$ with small p disturbs the transmission of a letter or information in small-world networks.  This is in striking contrast to the cases of scale free networks where an increase in $C(p)$ enhances the transmission. This means that the effect of $C(p)$ on transmission of information strongly depends on the basic topology of networks.  

 Thus $q$ greatly depends on the sum of $C(p) (0\leq p\leq q)$, which represents the state of the cycle structures up to $q$. This indicates that generalized clustering coefficients introduced in this article are effective indices to explore $q$-$th$ degrees of separation.

\section{SUMMARY }
\hspace{5mm}We attacked general $q$-$th$ degrees of separation based on the string formalism that is reformulated by adjacent matrix description, especially how cycle structures on networks, whose information is charged with generalized clustering coefficients, affect the separation number $q$. 

We see that the scale free network with $\gamma= 3.0$, which many real-world networks have about this value of exponent, just shows six degrees of separation and the four degrees of separation in networks with $\gamma <2.5$ but we do not recognize that nodes are linked together for $\gamma \geq 3.5$ up to six degrees of separation.

Six degrees of separation is mostly observed at small-world networks even with small $\alpha$ but nodes almost do not connect each other in 2 steps, even if $\alpha=1$ (random graph). This is consistent with the fact that the average path length rapidly decreases with $\alpha$ in small-world networks.

 There are a distinct difference between scale free networks 
and small-world networks in the relation between generalized clustering coefficients and $q$. We found that $q$ at scale free networks depends on generalized clustering coefficients more strongly than small-world networks. 

It would become an important issue how the basic topology of networks depends on the relation between generalized clustering coefficients and $q$ in more general cases.



\end{document}